# Title：

Single-shot fast 3D imaging through scattering media using structured illumination

# Author：

Aiping Zhai, Yuancheng Li, Wenjing Zhao, Dong Wang*

# Affiliations:

Key Laboratory of Advanced Transducers and Intelligent Control System, Shanxi Province and Ministry of Education, College of Physics and Optoelectronics, Taiyuan University of Technology, Taiyuan 030024, China.

*Corresponding author, Email: wangdong@tyut.edu.cn

# Abstract:

Conventional approaches for 3D imaging in or through scattering media are usually limited to 2D reconstruction of objects at some discontinuous locations, although the time-consuming iteration, guide-star, or complex system are implemented. How to quickly visualize dynamic 3D objects behind scattering media is still an open issue. Here, by using structured light illumination, we propose a single-shot technique that can quickly acquire continuous 3D surfaces of objects hidden behind the diffuser. The proposed method can realize the 3D imaging of single, multiple, and dynamic targets from the speckled structured light patterns under broad or narrow band light illumination, in which only once calibration of the imaging setup is needed before conducting the imaging. Our approach paves the way to quickly visualize dynamic objects behind scattering media in 3D and multispectral.

# 1. Introduction

Optical imaging is one of the most straightforward ways to express the physical world which is always three-dimensional (3D) around us. 3D imaging of an object hidden behind a scattering medium is crucial in various areas and applications such as biomedical imaging, deep tissue microscopy, colloidal imaging, astronomy, etc.[1-4]. It is challenging to acquire a clear image through these inhomogeneous samples since they induce light scattering. Several different approaches have been reported to remove the roadblock in 3D imaging inside or through scattering media, including 3D ballistic imaging which requires a priori knowledge of the target position to calibrate the gating image[5, 6]. However, it always becomes impractical for thick multiply scattering samples, because there are hardly any ballistic light components remaining. Considering the fundamental problem that the random wavefront introduced by the photon scattering when light impinged on the inhomogeneous samples, the wavefront shaping techniques based on optical phase conjugation (OPC), iterative algorithms, or transmission matrix measurement were put forward to "demodulate" the random wavefront of the scattering light[7-10]. Unfortunately, almost all of the iterative search, as well as transmission matrix measurement, are time-consuming and the conventional OPC needs to introduce a "guide-star" as the reference field for interference, resulting in the sensitivity and complexity of the imaging system.

Apart from these approaches above, taking advantage of the memory effect (ME) in disordered media [11], speckle patterns created by the interference of scattered light are always useful for image recovery[12], and then non-invasively single-shot 2D imaging through scattering media is demonstrated[13], which inspires researches on multi-spectral imaging[14], super-resolution imaging [15], multi-view imaging [16] and 3D imaging through scattering media[17-23]. In Ref.[19], based on the 3D ME of the scattering media, noninvasive tomographic imaging was developed by implementing 3D autocorrelation of axially distributed speckle images and a 3D phase retrieval algorithm. Instead of axially scanning the image sensor, approximately the same results can be obtained by scaling a single speckle image multiple times [20]. Besides, A multi-view imaging modality also enabled the depth-resolved imaging of objects hidden behind scattering media[21]. Phase-space imaging was proved available for distinguishing multiple point sources three-dimensionally localized inside scattering material[22, 23]. However, thus far, how to quickly visualize dynamic objects behind scattering media in 3D is still a challenging open issue.

Here, we propose a method using structured light illumination for single-shot fast 3D imaging through scattering media. The proposed method is not only to develop a fast 3D imaging technique for seeing through scattering media but also to try acquiring the 3D surface of the object which in general is non-planar and described continuously in 3D space. As shown in Fig. 1(a), by projecting structured light onto the object hidden behind the scattering media, a camera on the observer side captures a speckle pattern as shown in Fig. 1(c), the speckle pattern has the information of the reflected structured light modulated by the object's 3D surface, which is called the deformed fringe pattern. As a result, the high-quality reconstruction of the deformed fringe pattern from the speckle pattern, as well as the successful implementation of Fourier transform profilometry (FTP), allows single-shot fast 3D imaging through scattering media. As a proof of concept, we demonstrate single-shot broadband 3D imaging of a several-millimeter lengths DuPont line hidden behind the scattering media. Then, we demonstrate single-shot 3D imaging of the same DuPont line using a narrowband structured light illumination, and the 3D surface of the DuPont line can be better reconstructed, in other words, the quality of the 3D imaging can be improved. We also demonstrate single-shot 3D imaging of multiple objects hidden behind the scattering media. Finally, Single-shot fast 3D imaging of a rotating target was demonstrated. Using the proposed method, one can quickly visualize dynamic objects behind scattering media in 3D and multispectral.

## 2. Principle

Fig. 1 shows the schematic diagram of the proposed technique, while the detail of the experimental setup was provided in section 1, supplementary 1. Supposing that an object hidden behind scattering media was illuminated by a sinusoidal structured light, which will be deformed and reflected by the irregular 3D surface of the object. In this process, the 3D information of the object has been encoded in the deformed sinusoidal fringe pattern as shown in Fig. 1(b). At a conventional facility, it is very easy to obtain the 3D data from the deformed fringe pattern. However, when there is an optical diffuser stood between the camera and object, as shown in Fig. 1(a), the reflected light will get scattered, resulting in a speckle pattern formed on the photo-surface of the camera. Now, it is crucial to retrieve the deformed fringe pattern from the 2D speckle pattern for acquiring the 3D information of the object.

What made this process exciting is that the diffuser can be regarded as an imaging lens based on speckle intensity correlation in the ME region[24] and the image recorded by the camera can be considered as a convolution of the image of the hidden object within the ME and the point spread function (PSF) of the optical diffuser based imaging system which is complex, but in practice deterministic[25]. The recorded image, which is a speckle pattern as shown in Fig. 1(c), can be expressed as, $I = O * PSF$, where $*$ is the covolution operation and $O$ represents the deformed fringe pattern as shown in Fig. 1(b). Therefore, the image of the deformed fringe pattern accordingly can be recovered by deconvolution, $O' = deconv(I, PSF) \cong O$, as shown in Fig. 1(d) &(e). One can easily imagine that determining the PSF of the scattering imaging system in advance is of considerable importance, since the quality of it will affect the performance of the system. In the above reconstruction of the deformed fringe pattern, the PSF of the imaging system only needs to be calibrated once, the detailed implementation of the calibration can be found in section 2, supplementary 1.

How to acquire the 3D information of the object hidden behind scattering media from a single reconstructed deformed fringe pattern is an issue that FTP [26, 27] can be utilized to solve (see the details in section 3, supplementary 1). Based on the reconstructed deformed fringes pattern in Fig. 1(e), the true 3D information of the object can be reconstructed by FTP as shown in Fig. 1(f). Since FTP can do the 3D measurement of an object in real-time[28], thus single-shot fast 3D imaging through the scattering media can be achieved.

In brief, to realize single-shot fast 3D imaging through scattering media, we need to reconstruct the deformed fringe pattern from the speckle pattern first, and then implement FTP to acquire the 3D information of the object.

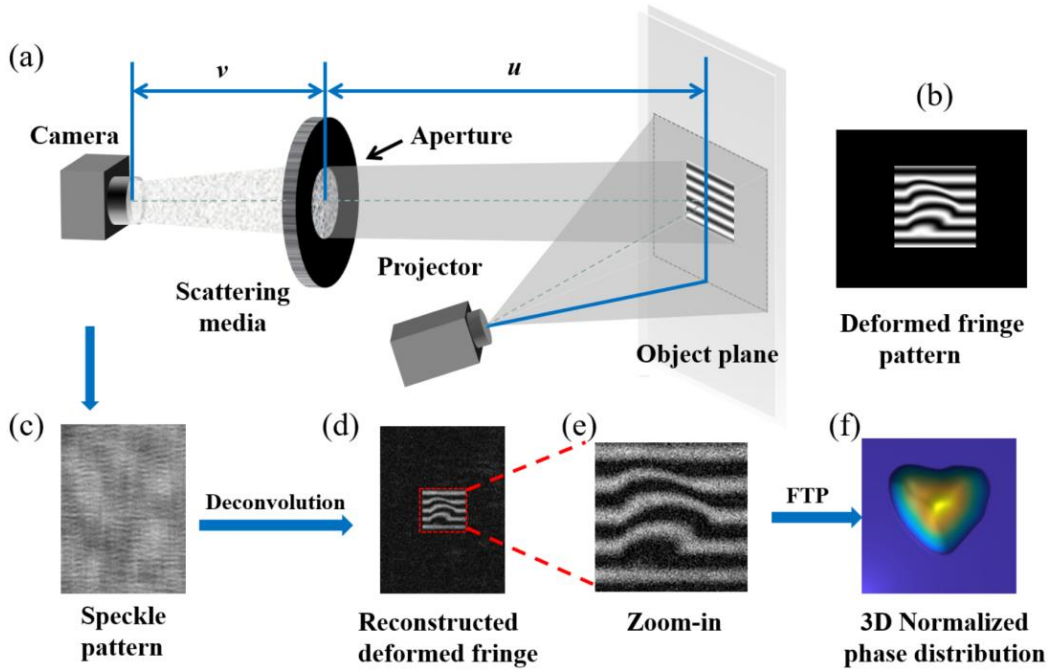

Fig. 1 Single-shot fast 3D imaging through scattering media. (a) Experiment setup. When a fringe pattern generated by PC is projected onto the object, it will be distorted as shown in (b) and reflected by the 3D surface of the object. The reflected light propagates through the diffuser and is captured by the camera as a speckle pattern shown in (c). (d) The deformed fringe is retrieved via deconvolution using the PSF of the system and speckle pattern (c). For clarity, its zoom-in is shown in (e). By Fourier transforming, filtering, and phase unwrapping, the 3D information of the object is acquired as shown in (f)

## 3. Experimental results

As a proof of concept, based upon the setup detailed in section 1, supplementary 1, an experimental demonstration of single-shot 3D broadband imaging of a target hidden behind scattering media is given in Fig. 2. A several-millimeter lengths DuPont line hidden behind the ground-glass diffuser is selected as the target object to be imaged, as shown in Fig. 2(a). By projecting a sinusoidal structured light onto the DuPont line, the structured illumination is thus modulated and reflected by the surface of the DuPont line. And then the reflected light propagates through the diffuser and gets scattered, giving rise to a noise-like speckle pattern on the photo-surface of the camera, as shown in Fig. 2(d). The noise-like speckle pattern has the information for retrieving the 3D object illuminated by the structured light. Within the ME region, the speckle pattern can be regarded as a convolution of the 2D image of the structured illumination modulated by the DuPont line and the system's PSF. Therefore, the 2D image of the structured illumination modulated by the DuPont line, i.e. the deformed fringe pattern can be retrieved by deconvolution as shown in Fig. 2(e). Using this deformed fringe pattern, the 3D information of the DuPont line can be obtained by FTP as shown in Fig. 2(f), of which the phase distribution at every point can be regarded as the function of position (x, y) in the Cartesian coordinates.

For comparison, we removed the diffuser and replaced it with an optical lens for imaging the deformed fringe pattern directly as shown in Fig. 2(b). And the 3D information of the DuPont line can be reconstructed by FTP and shown in Fig. 2(c). As can be seen from the comparison between Fig. 2(f) and Fig. 2(c), the proposed method can realize single-shot 3D broadband imaging of a single target object through scattering media, and the imaging quality is slightly degraded. The degradation is reasonable since the deformed fringe pattern retrieved in Fig. 2(e) is a little bit blurry. Further, we believe that the reason for the blurry deformed fringe pattern may be mainly because the illumination light source is broadband, light with different wavelengths gets scattered by the diffuser and thus enhanced mutual interferences, which makes the reconstruction of the deformed fringe pattern blurry.

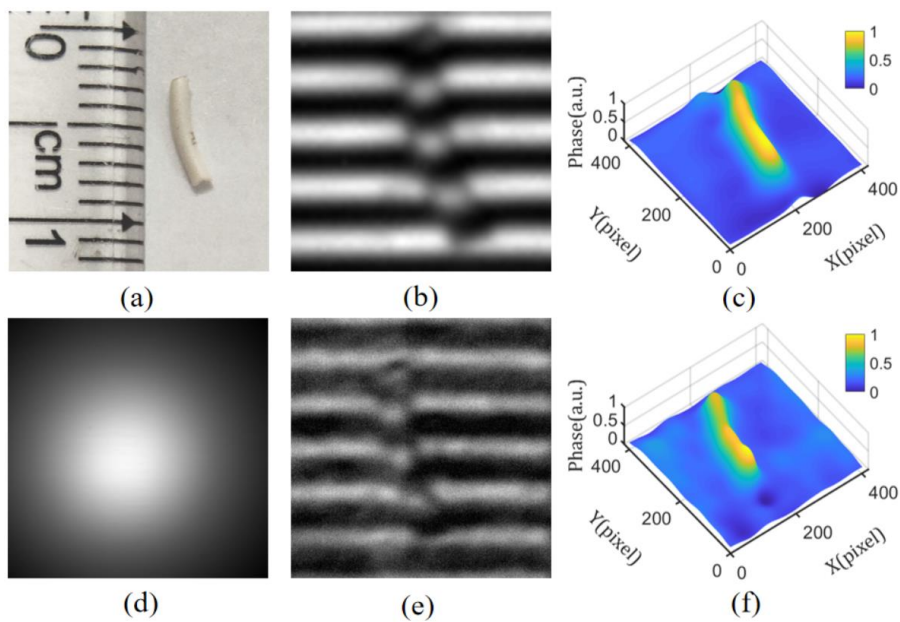

Fig. 2 Single-shot 3D imaging of a single object through scattering media using broadband structured light illumination. (a) The object to be imaged. (b) the deformed fringe pattern captured by the camera with an optical lens. (c) 3D information of the object is obtained by performing FTP on (b). (d) The speckle pattern is captured by the camera through a single shot exposure after the deformed structured light is reflected and passed through the diffuser. (e) The deformed fringe pattern is retrieved from deconvolution using the PSF and speckle pattern (d). (f) The 3D information of the object is obtained by performing FTP on (e).

To verify the above-mentioned inference and improve the reconstruction accuracy of the proposed method, a 10 nm narrowband light filter with a center wavelength of 635 nm was added to the setup and a narrowband structured light was obtained as shown in Fig. S2 (in section 1, supplementary 1). Fig. 3 gives the experimental results, when the same object as shown in Fig. 2(a), hidden behind the diffuser, was illuminated by the narrowband sinusoidal structured light. By projecting the sinusoidal structured light onto the DuPont line, the structured illumination is thus modulated and reflected by the surface of the DuPont line. And then, the reflected light propagates through the diffuser and gets scattered. The corresponding noise-like speckle pattern, captured by the camera through a single shot exposure, is shown in Fig. 3(a). Here, we measured the system PSF in advance with the narrowband light illumination (Fig. S4, Supplementary 1). Fig. 3(b) shows the deformed fringe pattern, which was obtained by deconvolution using the speckle pattern and the narrowband PSF. The 3D information of the DuPont line was reconstructed by FTP as shown in Fig. 3(c). Obviously, the 3D information of the DuPont line reconstructed using the narrowband illumination is better than that reconstructed using broadband illumination. This is reasonable since here the deformed fringe pattern, retrieved and shown in Fig. 3(b), exhibits a better sinusoidal distribution than that in Fig. 2(e). The results justify our preceding inference that the reason for the blurry deformed fringe pattern in Fig. 2(e) is mainly because the illumination light source is broadband, light with different wavelengths gets scattered by the diffuser and thus enhanced mutual interferences, which makes the reconstruction of the deformed fringe pattern blurry. This also suggests that using narrowband illumination in the proposed method can improve the reconstruction accuracy of the 3D object hidden behind scattering media.

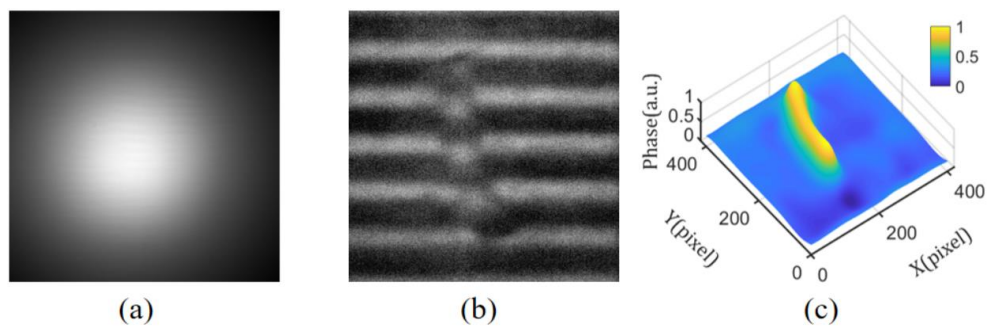

Fig. 3 Single-shot 3D imaging of a single object through scattering media using the narrowband structured light illumination. (a) The speckle pattern is captured by the camera through a single shot exposure. (b) The deformed fringe pattern is reconstructed by deconvolution of PSF and speckle pattern. (c) The 3D information of the object is obtained by performing FTP on (b).

Besides the single-shot fast 3D imaging of a single target, we try to reconstruct multiple objects hidden behind the scattering media. In this case, for higher accuracy, narrowband structure light generated in the same way above was adopted as the illumination source. Fig. 4(a) shows the two objects to be imaged, which were placed next to each other behind the ground glass diffuser. By conducting the same fringe projection and measurement as the previous experiments do, the camera captured the signal through a single shot exposure, which is still the noise-like speckle pattern as shown in Fig. 4(b). As shown in Fig. 4(c), the deformed fringe pattern can be obtained by conducting deconvolution using the speckle pattern in Fig. 4(b) and the narrowband PSF. The 3D information of the two objects can be reconstructed by FTP as shown in Fig. 4(d). The results suggest that the proposed method can realize single-shot 3D imaging of multiple objects through scattering media.

However, as compared to the result in Fig. 3(c), the reconstruction quality of the DuPont line is slightly degraded, and this is mainly because the object was placed in a different orientation, and thus the illumination angle is different although using the same sinusoidal structured light illumination, which is an issue of FTP (see the details in section 3, supplementary 1). The same phenomena can be also found in Fig. 5 and the corresponding Visualization 1 in supplementary 2.

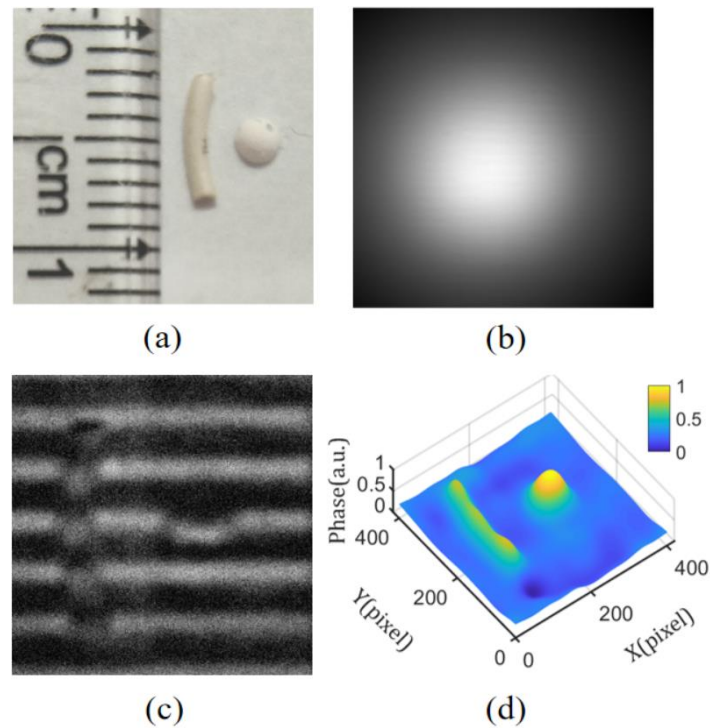

Fig. 4 Single-shot 3D imaging of multiple objects through scattering media. (a) The two objects. (b) The speckle pattern is captured by a single shot of the camera. (c)The deformed fringe pattern is retrieved from deconvolution using the PSF and speckle pattern (b). (d) The 3D information of the two objects is obtained by performing FTP on (c).

As can be noted from the above experimental demonstrations, by projecting a broad or narrow band sinusoidal fringe pattern onto the objects hidden behind scattering media, the 3D information of not only a single object but also multiple objects can be quickly reconstructed from a single shot speckle pattern captured by the camera. The proposed method is expected to realize dynamic 3D imaging through scattering media. To demonstrate this, a 0.7 mm width, 0.45 mm height, and 5.6 mm length rectangular iron sheet, driven by a dynamically rotating magnet, was selected as the dynamic object hidden behind scattering media. By projecting the narrowband sinusoidal structured light to illuminate the rotating rectangular iron sheet, the camera captured a series of speckle patterns at an interval of 2 s which was limited by the single-shot exposure time of the camera for getting enough SNR. Part of the captured speckle patterns is given in Fig. 5(a). The corresponding deformed fringe patterns can be retrieved quickly by performing deconvolution using the speckle patterns and the pre-recorded PSF as shown in Fig. 5(b). For each frame of the deformed fringe patterns reconstructed, it only took about 27 ms. Using the deformed fringe patterns, the 3D information of the rotating rectangular iron sheet can be reconstructed by FTP as described in Fig.

5(d), and the top views of them are given in 5(c). For each frame of the reconstructed 3D information of the rotating rectangular iron sheet by FTP, it only took around 38 ms. The results suggest that the proposed method can realize fast 3D imaging through scattering media, and the imaging speed is mainly limited by the single-shot exposure time of the camera for capturing a frame of speckle patterns, which is 2s since here the narrowband illumination was used. The single-shot exposure time of the camera for capturing one frame speckle pattern was 350ms under the condition of broadband illumination with approximately 150 lumens at 15 W LED power consumption (see section 1, Supplementary 1). The imaging speed of the proposed technique can be further improved to be approached in real-time by increasing the power of the light source and using a more sensitive camera, e.g. the more powerful EMCCD.

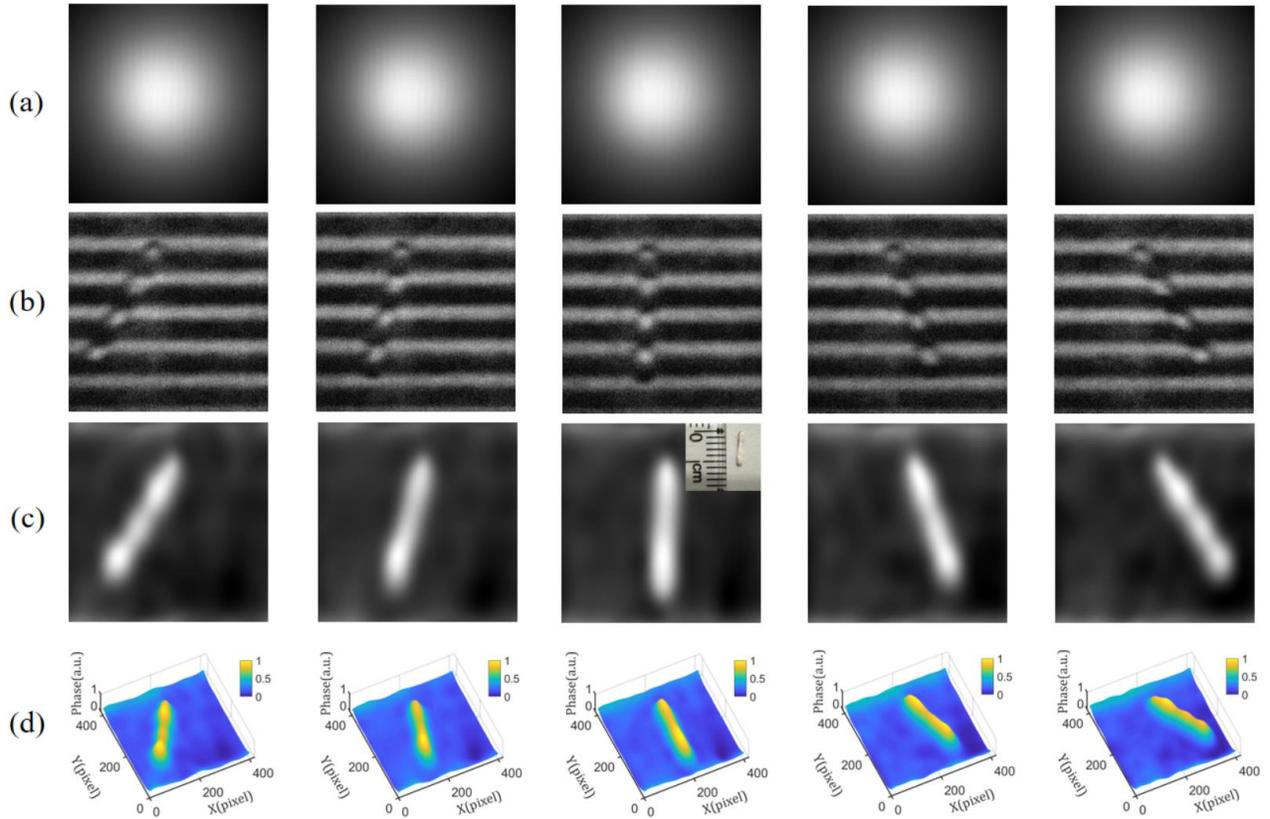

Fig. 5 Single-shot fast 3D imaging through scattering media. (a) Part of the speckle patterns, which are captured by the camera when the object is rotating. (b) The deformed fringe patterns, which are retrieved from the speckle patterns. (c) Top views of the object at different times. (d) 3D displays of the rotating object. More details can be seen in Visualization 1.

## 4. Discussion and conclusion

We have successfully demonstrated a method using structured light illumination for single-shot fast 3D imaging through scattering media. However, the proposed method could be further improved.

First, in this work, the structured fringe pattern was assumed to project directly onto the object. In practice, if the object was hidden inside scattering media, how to realize an available structure light illumination becomes a key factor for the successful implementation of the proposed 3D imaging technique. The achievements on structured light shaping and controlling based on machine learning [29, 30] or wavefront shaping [31-33] may be useful for solving this problem.

Second, since the ME-based deconvolution was used to retrieve the deformed fringe pattern from the speckle pattern, conventionally, the field of view (FOV) and depth of field (DOF), which directly determine the imaging scope of the 3D object, are respectively restricted in the lateral and axial ME region of scattering media[34]. Besides, the imaging quality will be degrading along with the shrunken ME region if the object was hidden behind stronger scattering media, which has been demonstrated in Fig. S7 of section 4, Supplementary 1. Nevertheless, it should be noted that although the deformed fringe pattern was retrieved using the ME-based deconvolution scheme in the method, this is not the only choice, since the transmission matrix measurement[35] and time-reversal optical phase conjugation [36] could be also utilized for retrieving the deformed fringe pattern. In addition, methods dedicated to imaging beyond the ME region can be beneficial for enlarging the FOV or DOF [37-40]. For example, an object beyond the lateral ME region can be reconstructed by utilizing correlations among the regional PSFs to reconstruct various parts of the object and then stitch the multiple parts to be the whole object.

Third, to pursue the fast imaging speed, FTP was used to acquire the 3D information of the object from the deformed fringe pattern, in which spectrum leakages were inevitable when the fundamental frequency was filtered in the frequency domain to obtain the wrap-phase information from the deformed fringe pattern, resulting in some details of the object was missing. The phenomenon can be noted in Fig. 4(d). Fortunately, N-step phase shift technology can obtain more accurate wrap-phase information directly from the intensities in the N fringe patterns through anti-trigonometric operation, but the imaging speed will be decreased by 1/N[41, 42]. The N-step phase shift technology can be adopted in the method for higher-quality 3D imaging of the object hidden behind scattering media.

Fourth, at the present stage, the PSF of the scattering-media-based imaging system needs to be calibrated once in advance before imaging an object, and once scattering media changes the PSF needs to recalibrate again. This limits the method to be used for imaging through dynamic scattering media. However, if one can estimate the PSF from the captured speckle pattern, the limitation can be removed. Inspiringly, Ref.[15] has recently demonstrated an ingenious technique for non-invasive super-resolution imaging through dynamic scattering media, in which the PSF of the scattering-media-based imaging system can be estimated from the captured speckle pattern.

In summary, we have demonstrated single-shot fast 3D imaging through scattering media using structured illumination, in which by projecting a broad or narrow band sinusoidal structured illumination onto objects hidden behind scattering media, the 3D information of them can be quickly reconstructed from a single shot speckle pattern captured by the camera. The experiments with broadband light illumination suggest the method can realize single-shot 3D broadband imaging through scattering media, and this implies it has the potential for multispectral 3D imaging through scattering media, in which only the calibrations of PSFs with different bandwidth wavelengths are demanded. The experiments with narrowband light illumination suggest using a narrowband light source can improve the reconstruction accuracy of the proposed method. We have also demonstrated it can do fast 3D imaging of a dynamic object through scattering media, and the imaging speed is mainly limited by the single-shot exposure time of the camera for capturing a frame of speckle patterns. The imaging speed of the method could be further improved to be approached in real-time by increasing the power of the light source and/or using a more sensitive camera e.g. the more powerful EMCCD. Our method paves the way to quickly visualize dynamic objects behind scattering media in 3D and multispectral.

**Funding.**

National Natural Science Foundation of China (61805167); Research Project Supported by Shanxi Scholarship Council of China (SSCC2021).

**Acknowledgments.**

We thank the support of the National Natural Science Foundation of China and the support of the Shanxi Scholarship Council of China.

**Disclosures**

The authors declare no conflicts of interest.

# Supplementary document

**Title：**

Single-shot fast 3D imaging through scattering media using structured illumination

**Author：**

Aiping Zhai, Yuancheng Li, Wenjing Zhao, Dong Wang*

**Affiliations:**

Key Laboratory of Advanced Transducers and Intelligent Control System, Shanxi Province and Ministry of Education, College of Physics and Optoelectronics, Taiyuan University of Technology, Taiyuan 030024, China.

*Corresponding author, Email: wangdong@tyut.edu.cn

## 1. Experiment detail：

The schematic diagram of the experimental setup is displayed in Fig. S1. Broadband (420 nm~700 nm), incoherent structured light was projected onto the object by a digital projector (DLP Lightcrafter 4500 with a resolution of 912×1140 pixels). After impinging on the surface of the object, the structured light was reflected and passed through the optical diffuser (LBTEK, DW110-1500), resulting in a speckle pattern on the photo-surface of the camera (Andor Zyla 4.2, sCMOS with resolution of 2048×2048 pixels, pixel size 6.5×6.5 μm), controlled by the Matlab code running on a computer with an Intel Core i9-10900K CPU @ 3.7GHz and a graphics card of NVIDIA GeForce RTX3080. In this case, 3D imaging through scattering media was demonstrated in a field of view (FOV) of 10 mm × 10 mm, that is the illumination area of the structured light as shown in Fig. S1. Since the scattering media in the system can be regarded as a 'scattering imaging lens' within the memory effect (ME) region. Therefore, distance from the optical diffuser to the object plane and the image plane can be considered as object distance denoted as $u$ and image distance denoted as $v$ in Fig. S1 respectively. To image within the ME region and satisfy the sampling theorem, $u$, $v$, and the aperture size were set to 250 mm, 80 mm, and 4.5 mm respectively. The object to be imaged was fixed in the white region of a flat resin board, on which black cardboard was glued as the background for reducing reflections out of the target imaging range. Normals of the camera photo-surface, optical diffuser, and the object plane were overlapped with the optical axis. Structure light illuminated the object at an angle of $\theta \approx 20°$ to the system optical axis with approximately 150 lumens at 15 W LED power consumption on the condition of broadband illumination.

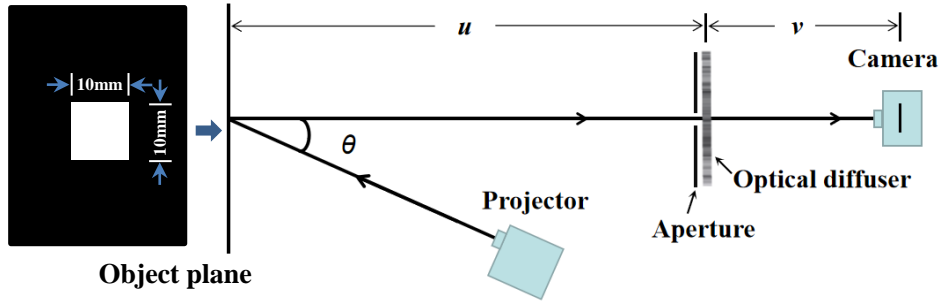

Fig. S1 Schematic diagram of the experimental setup using broadband structured light illumination

The PSF measured by capturing the speckle pattern corresponding to a broadband point source may be blurred, and this is mainly because the illumination light source is broadband, light with different wavelengths gets scattered by the diffuser and thus enhanced mutual interferences, which makes the recorded PSF blurry. To verify the performance in the condition of narrowband structured light illumination, the setup was thus further modified as shown in Fig. S2, in which a filter with the center wavelength of 635 nm was placed in front of the projector.

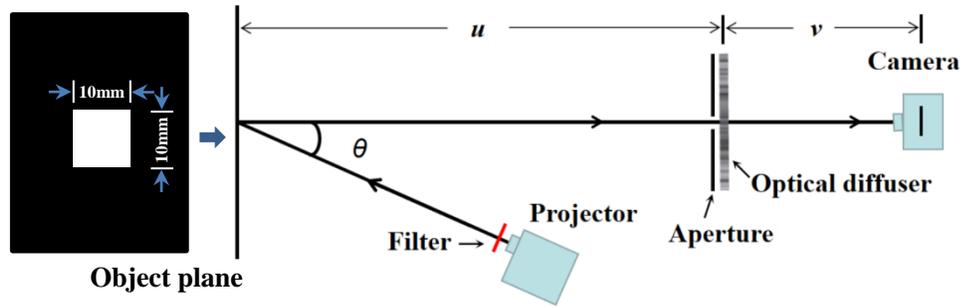

Fig. S2 Schematic diagram of the experimental setup using narrowband structured light illumination

## 2. PSF measurement

Before implementing deconvolution, PSF calibration of the scattering imaging system should be accomplished in advance. For a determined system, the PSF calibration only needs to be done once, while the quality of the PSF is very important to the final imaging result. By projecting a point source on the center of the object plane as shown in Fig. S1, the light emitted from the point passes through the optical diffuser and forms a speckle pattern (i.e. PSF of the scattering imaging system), which was recorded by the camera with its maximum exposure time of 30 s.

In the experiments, point sources with the sizes of $1\times1$ pixel, $2\times2$ pixels, $3\times3$ pixels, $4\times4$ pixels, and $5\times5$ pixels were projected onto the object plane. And the speckle patterns, as shown in Fig. S3, were collected by the camera corresponding to each point source. Theoretically, the smaller the point source is, the higher the imaging quality will be. However, the low-light detection capability of the camera limits its performance, and a long time exposure will result in high dark noise and low signal-to-noise ratio ( SNR ). As a result, as shown in Fig. S3, the contrast of the speckle pattern corresponding to the $1\times1$ pixel point source is lower than that of other speckles. For selecting a better system PSF, we try to reconstruct a sinusoidal fringe pattern hidden behind scattering media by implementing deconvolution using the PSFs in Fig. S3 (a) and the speckle pattern corresponding to the sinusoidal fringe pattern. The reconstructed fringe patterns are displayed in Fig. S3 (b), from which we can see the contrasts of the first two fringe patterns reconstructed with the PSFs that were measured by 1×1 pixel and 2×2 pixels point source projecting are relatively lower than that of other fringe patterns. Considering the granularity of the speckle pattern and the sinusoidal property of the reconstructed fringe pattern, the speckle pattern corresponding to the 3×3 pixels point source was selected as the PSF of the system.

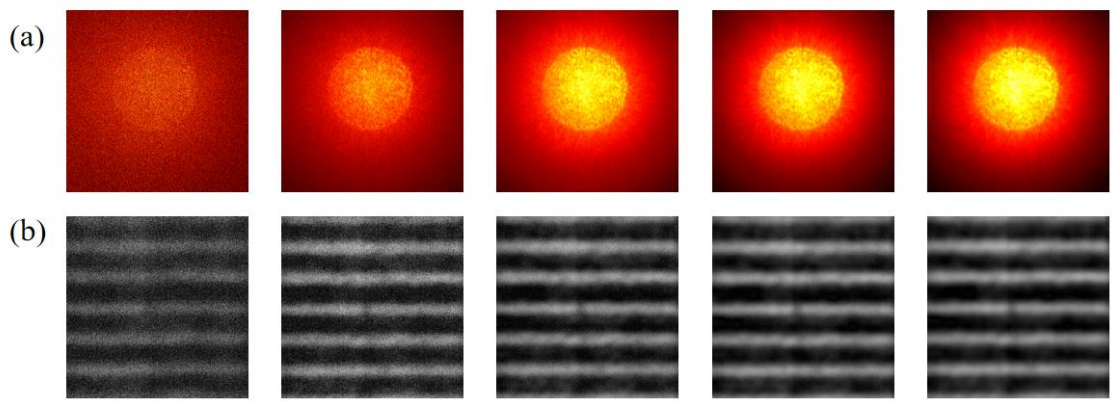

Fig. S3 PSF measurement using broadband structured light illumination (a) Speckle patterns corresponding to the broadband point sources from left to right with the sizes of 1×1 pixel, 2×2 pixels, 3×3 pixels, 4×4 pixels, and 5×5 pixels. (b) Reconstructed fringe patterns by implementing deconvolutions using the system PSFs and the speckle patterns.

PSF measurement using the narrowband structured light illumination was the same as above. In this case, the quality of the fringe pattern reconstructed using the PSF corresponding to the narrowband 1×1 pixel point source was slightly worse than that with the broadband 1×1 pixel point source illumination due to the light intensity attenuation caused by light filtering.

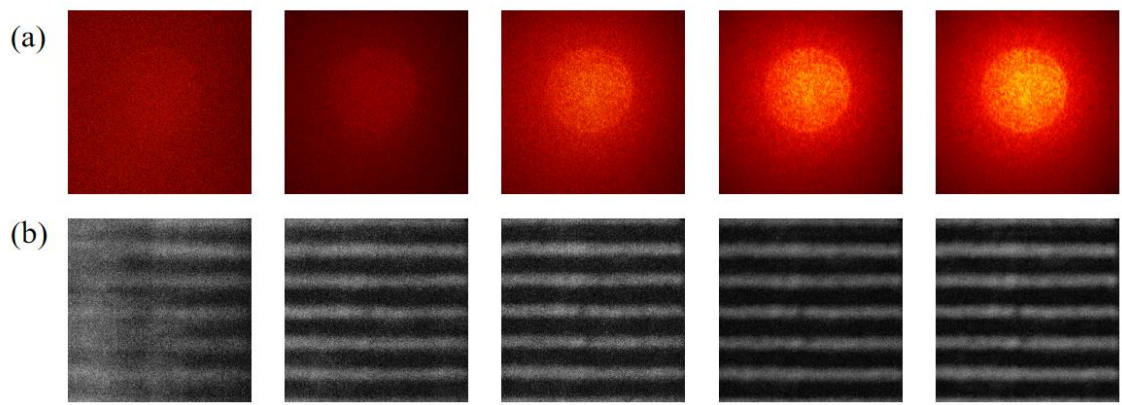

Fig. S4 PSF measurement using narrowband structured light illumination: (a) Speckle patterns corresponding to the narrowband point sources from left to right with the sizes of 1×1 pixel, 2×2 pixels, 3×3 pixels, 4×4 pixels, and 5×5 pixels. (b) Reconstructed fringe patterns by implementing deconvolutions using the system PSFs and the speckle patterns.

## 3. Fourier transform profilometry

The single-shot nature of FTP makes it highly suitable for the 3D shape measurement of dynamic surfaces [1]. The phase is extracted by applying a properly designed band-pass filter in the frequency domain. FTP for 3D shape measurement is usually implemented as the following.

Commonly, the general geometry is shown in Fig. S5, in which the optical axes $P_1P_2$ of a projector lens crosses that of camera lens $I_1I_2$ at point O on a reference plane R, which is a fictitious plane normal to $I_1I_2$ and serves as a reference, from which object's 3D phase distribution can be measured.

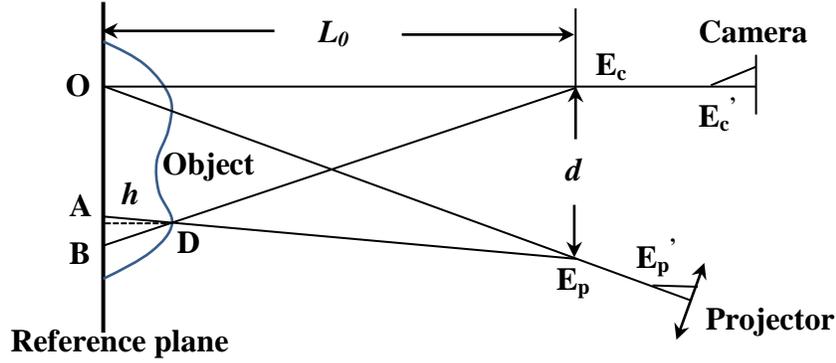

Fig. S5 Optical geometry for FTP

In the imaging system shown in Fig. S5, when a sinusoidal structured light is projected onto an object, which was placed on the reference plane in the FOV of 10 mm × 10 mm, the deformed fringe pattern, modulated by the object's surface, can be expressed as,

$$g(x, y) = r(x, y) \sum_{n=-\infty}^{\infty} A_n \exp\{i[2\pi n f_0 x + n\varphi(x, y)]\} \tag{S1}$$

when $h(x, y) = 0$, which means the sinusoidal fringe pattern was projected onto the reference plane, the fringe pattern captured by the camera can be written as,

$$g_0(x, y) = r_0(x, y) \sum_{n=-\infty}^{\infty} A_n \exp\{i[2\pi n f_0 x + n\varphi_0(x, y)]\} \tag{S2}$$

where $r_0(x, y)$ and $r(x, y)$ are respectively the non-uniform distributions of reflection on the diffuse object and the reference plane. $A_n$ is the weighting factors of Fourier series, $f_0$ is the fundamental frequency of the observed fringe pattern, $\varphi(x, y)$ and $\varphi_0(x, y)$ are the phase modulation resulting from the object surface and original phase modulation on the reference plane, respectively.

Compute the 2-D Fourier transform of Eq. (S1) and Eq. (S2) and obtain the Fourier spectra in the frequency domain. A suitable filter is important to acquire an accurate 3D surface reconstruction [2]. In this case, the Hanning filter window was used to select the fundamental component for 3D shape recovering. Therefore, in the process of FTP, both the reconstruction accuracy and the maximum range are dependent on the degree of the spectrum aliasing and the filtering effect. With inverse Fourier transform, a complex signal modulated by the 3D surface of the object was obtained as follow:

$$\hat{g}(x, y) = A_1 r(x, y) \exp\{i[2\pi n f_0 x + \varphi(x, y)]\} \tag{S3}$$

The same operation applied on the fringe pattern can be done when the sinusoidal fringe pattern was projected on the reference plane,

$$\hat{g}_0(x, y) = A_1 r_0(x, y) \exp\{i[2\pi n f_0 x + \varphi_0(x, y)]\} \tag{S4}$$

Ignoring all the geometry errors, the unwrapped phase distribution related to the height distribution of the object's 3D surface can be obtained:

$$\Delta\varphi(x, y) = \varphi(x, y) - \varphi_0(x, y) = 2\pi f_0 \overline{AB} = \arctan \frac{\text{Im}\left[\hat{g}(x, y) * \hat{g}_0^*(x, y)\right]}{\text{Re}\left[\hat{g}(x, y) * \hat{g}_0^*(x, y)\right]} \tag{S5}$$

where *Im* and *Re* represent the imaginary part and real part of $\hat{g}(x, y) * \hat{g}_0^*(x, y)$ respectively. From Eq. (S5), a series of discontinuous phases are distributed in the range of [-π, π], which is the period of the deformed fringe pattern. The continuous phase can be obtained by various phase unwrapping algorithms [3-5], and in this work, the quality-guided spatial phase unwrapping algorithm [3] was used in the phase unwrapping process.

In this work, before implementing the FTP, the deformed fringe patterns, corresponded to the sinusoidal fringe patterns projected on the object and the reference flat plane, must be recovered from the speckle pattern. Fig. S6 (a) gives the recorded speckle pattern when the structured light illuminates the reference flat plane. Fig. S6 (b) gives the fringes pattern on the reference plane, which was retrieved through deconvolution of the narrowband PSF and the speckle pattern. The unwrapped phase as shown in Fig. S6 (f) was obtained by implementing the Fourier transform, filtering, calculation, and phase unwrapping. The same process can be performed when an object was placed on the reference plane, resulting in the phase distribution related to the object's 3D surface can be acquired.

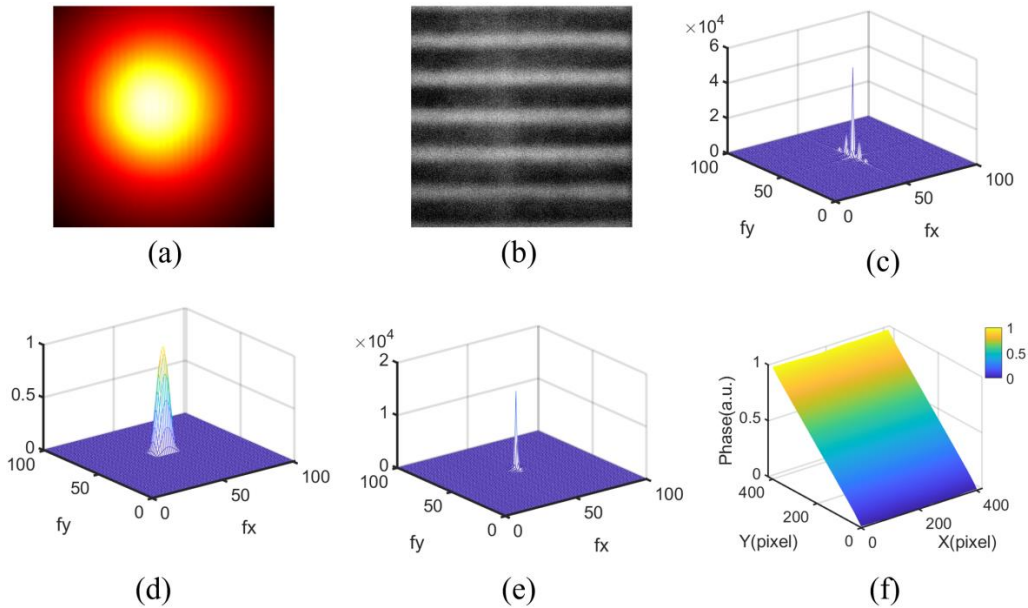

Fig. S6 Phase retrieval of the reference plane from the speckle pattern by the deconvolution and FTP: (a) The captured speckle pattern. (b) The reconstructed fringes pattern by deconvolution. (c) Fourier transform of the fringes pattern in (a). (d) The Hanning window. (e) Fundamental frequency obtained by performing Hanning window on (b). (f) The unwrapped 3D phase distribution of the reference plane.

It is worth noting that in structured-light 3D imaging using FTP, the reconstruction accuracy for an asymmetric object was influenced by the fringe direction [7] because the optimum fringe angle always brings the largest phase changes for a given surface of the object. Additionally, a suitable filtering window should be also selected for the different fringe directions to keep the image quality.

### 4. 3D imaging through stronger scattering media

One of the other things someone may be looking at, and again we are interested in, is what about the performance of this method when the scattering is stronger. Experiments with higher scattering media (LBTEK, DW110-600) were thus performed and the results were displayed in Fig. S7 of supplementary 1, from which we can see the image quality was slightly degraded a little bit, which is reasonable since the stronger scattering should make the ME region shrunken a little bit.

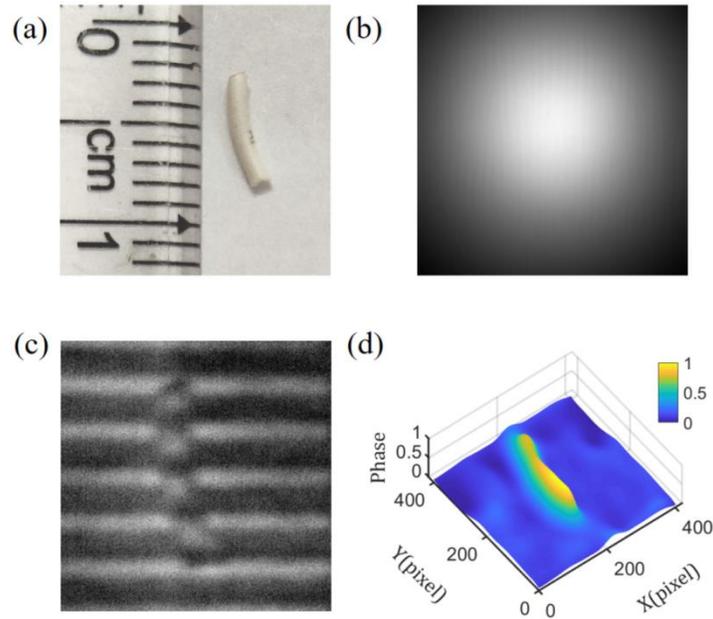

Fig. S7 3D imaging of a single object through a stronger scattering medium (LBTEK, DW110-600 ground glass diffuser) using narrowband structured light illumination. (a) The object to be imaged. (b) The captured speckle pattern. (c) The deformed fringe pattern, which is retrieved from the speckle pattern. (d) The 3D information, which is obtained by performing FTP on (c).